\documentclass[10pt]{aastex}
\begin{document}

\newcommand{\ms}{$M_{\odot}$}
\newcommand{\msb}{$M_{\odot}$~}
\newcommand{\al}{$^{26}$Al}
\newcommand{\alb}{$^{26}$Al~}
\newcommand{\fe}{$^{60}$Fe}
\newcommand{\be}{$^{10}$Be}
\newcommand{\ca}{$^{41}$Ca}
\newcommand{\mn}{$^{53}$Mn}
\newcommand{\pd}{$^{107}$Pd}
\newcommand{\tc}{$^{99}$Tc}
\newcommand{\pu}{$^{244}$Pu}
\newcommand{\hf}{$^{182}$Hf}
\newcommand{\ct}{$^{13}$C}
\newcommand{\cm}{$^{247}$Cm}
\newcommand{\tho}{$^{230}$Th}
\newcommand{\nb}{$^{92}$Nb}
\newcommand{\msun}{\ensuremath{M_\odot}}
\newcommand{\mdot}{\ensuremath{\dot{M}}}
\newcommand{\cd}{$^{12}$C}
\newcommand{\cdb}{$^{12}$C~}
\newcommand{\ctb}{$^{13}$C~}
\newcommand{\ctanb}{$^{13}$C($\alpha$,n)$^{16}$O~}
\newcommand{\ctan}{$^{13}$C($\alpha$,n)$^{16}$O}
\newcommand{\neanb}{$^{22}$Ne($\alpha$,n)$^{25}$Mg~}
\newcommand{\nean}{$^{22}$Ne($\alpha$,n)$^{25}$Mg}

\title{\large News on the s process from Young Open Clusters}

\author{Enrico Maiorca\altaffilmark{1,2,3}, Laura Magrini\altaffilmark{2},
Maurizio Busso\altaffilmark{1,3}, Sofia Randich\altaffilmark{2},
Sara Palmerini\altaffilmark{1,3}, Oscar Trippella\altaffilmark{1,4}}

\altaffiltext{1}{Department of Physics, University of Perugia, Via
Pascoli, 06122, Perugia, Italy;  emaiorca@gmail.com}

\altaffiltext{2}{INAF, Osservatorio Astrofisico di Arcetri, Largo
Enrico Fermi 5, 50125, Firenze, Italy}

\altaffiltext{3}{INFN, Section of Perugia, Via Pascoli, 06122,
Perugia, Italy}

\altaffiltext{4}{INFN, Laboratori Nazionali del Sud, Via Santa
Sofia, Catania}

\begin{abstract}
Recent spectroscopic measurements in open clusters younger than the
Sun, with [Fe/H] $\gtrsim$ 0, showed that the abundances of
neutron-rich elements have continued to increase in the Galaxy after
the formation of the Sun, roughly maintaining a solar-like
distribution. Such a trend requires neutron fluences larger than
those so far assumed, as these last would have too few neutrons per
iron seed. We suggest that the observed enhancements can be produced
by nucleosynthesis in AGB stars of low mass ($M <$ 1.5\ms), if they
release neutrons from the \ctanb reaction in reservoirs larger by a
factor of 4 than assumed in more massive AGBs ($M>$ 1.5\ms).
Adopting such a stronger neutron source as a contributor to the
abundances at the time of formation of the Sun, we show that this
affects also the solar s-process distribution, so that its {\it main
component} is well reproduced, without the need of assuming ad-hoc
primary sources for the synthesis of s elements up to $A \sim$ 130,
contrary to suggestions from other works. The changes in the
expected abundances that we find are primarily due to the following
reasons. i) Enhancing the neutron source increases the efficiency of
the s process, so that the ensuing stellar yields now mimic the
solar distribution at a metallicity higher than before ([Fe/H
]$\gtrsim -0.1$). ii) The age-metallicity relation is rather flat
for several Gyr in that metallicity regime, so that those conditions
remain stable and the enhanced nuclear yields, which are necessary
to maintain a solar-like abundance pattern, can dominate the
composition of the interstellar medium from which subsequent stars
are formed.

\end{abstract}


\section{Introduction}
Very recently, measurements of s-process elements in young Galactic
stellar systems \citep[][hereafter Paper I]{dor09,jac11,m+11}
indicated that the neutron-rich nuclei Y, Zr, Ba, La and Ce are
enhanced as compared to the Sun. The enhancement factor is $\simeq$
0.2 dex, and is the same (within uncertainties) for the heavier s
elements (Ba, La Ce, hereafter $hs$) and for the lighter ones (Y,
Zr, hereafter $ls$). Hence, a roughly solar ratio between the two
groups is maintained\footnote{The abundance of Sr, in general
difficult to obtain, could not be measured because its useful lines
were outside the wavelength range of the available spectra}.

The stellar systems involved are open clusters (hereafter OCs)
younger than the Sun, belonging to the recent evolutionary stages of
the Galactic thin disk. In the evolution of the disk, over several
Gyr, the observed metallicity [Fe/H] ($\equiv \log$ ($N$(Fe)/$N$(H))
$-$ $\log$ ($N$(Fe)/$N$(H))$_{\odot}$) does not increase appreciably
with time (rather, for every epoch the observations of [Fe/H] show a
considerable scatter). These facts hamper the possibility of
identifying definite trends in the abundances of heavy elements as a
function of iron \citep{mash1,mash2}. In the case of open clusters,
however, one has a reasonably good knowledge of their ages
\citep[see e.g.][]{msrg}, so that using this information we could
show the abundances of neutron-rich elements directly as a function
of time. This was the key to reveal clearly the mentioned
enhancement. \citep[Previous studies on field stars had already
tentatively suggested this as a potentially effective procedure,
see][]{mash1}.

Chemical and chemo-dynamical models of Galactic evolution
\citep{t+99,t+04,r+99}, based on the currently-accepted scenario for
s processing, predict a plateau, or even a decrease in the
abundances [$X_{i}$/Fe] of neutron-rich elements after the solar
formation; a late increase is certainly not expected.

The new results can be understood in either of two possible cases.
The first possibility lays in the late emergence of a purely
secondary-like neutron-capture process, i.e. one driven by a neutron
source not synthesized by the star from pure H, but derived from
elements inherited at stellar birth. The abundances of s elements
thus produced would grow steadily in time as compared to iron,
becoming appreciable only in advanced epochs. This would be the case
for the \neanb source, activated in Intermediate Mass Stars (IMS).
$^{22}$Ne is indeed produced by two $\alpha$ captures on $^{14}$N,
which in its turn derives from p-captures on the original CNO nuclei
collected from the interstellar medium (ISM). However, the present
knowledge of s processing in the Sun and in the Galaxy, and
especially of the low neutron densities at which it must occur
\citep{a+99}, excludes this possibility. Indeed, it would imply an
isotope mix in the solar composition, for elements sensitive to
reaction branchings along the s-process path, incompatible with
meteoritic abundances \citep[see e.g.][]{k+90,a+99}. It would also
produce an insufficient number of neutrons to feed effectively the
$hs$ elements \citep{b+88} and would fail to explain the observed
abundance ratios among Sr, Y and Zr in carbon stars \citep{a+01}.

The only alternative is that the contribution from the \ctanb
neutron source in low mass stars be larger than so far adopted in
AGB models. In such a case, however, it is important to verify that
any model change required to fit the OC data does not affect the
many constraints on AGB nucleosynthesis that have already been
accounted for \citep[see e.g.][]{b+99,b+01}. In the present paper we
aim at demonstrating that, for explaining the new observational
evidence, it suffices that the \ctb reservoir be larger than
previously assumed only in very low mass stars ($M < 1.5$\ms). This
would not modify previous interpretations of s-enriched AGB stars in
the solar neighborhoods, which have a higher mass \citep[about 1.5
to 2.5\ms, see e.g.][]{a+02,g+06,gb08}. It must be noted that the
standard models used so far for them (as well as for the Sun)
underproduce the ''light'' elements Sr, Y and Zr with respect to the
solar abundances \citep[see][for details]{t+04}. We instead aim at
showing that our assumptions have the further advantage of avoiding
this problem, thus accounting completely for the solar {\it main
component}, i.e. for the abundances of all s elements heavier than
Kr-Rb.

This paper is organized as follows. In Section 2 we briefly discuss
the changes necessary in the models in order to explain the new
observational data and we outline our approach. In Section 3 we
discuss the new stellar yields. In Section 4 we present our main
results, and substantiate the claim that they also explain the solar
s-process distribution. Then, in Section 5 we summarize our main
conclusions.

\section{Revised models for the s-element enrichment}
In the current literature, one usually finds the same assumptions on
the formation of the \ctb neutron source, for all LMS between 1 and
3\ms. In particular, the mass of the layer where \ctb is formed (the
so-called {\it \ct-pocket}) is kept constant in most s-process
nucleosynthesis models, while the average abundance by mass of \ctb
inside the pocket is parameterized, to give sufficient neutrons to
reproduce the observed constraints. \citep[One has actually to
recall that there is no ab-initio physical model for the formation
of this neutron source, but the observational data require s
processing from its activation in AGB stars, as discussed, e.g.,
by][]{g+98}.

Recently, looking for a better physically-justified scenario, the
proton penetration into the intershell zone was assumed to be the
consequence of a naturally-decaying velocity profile of the
convective eddies at the envelope border \citep{st06,st09,cr9}. This
decay was regulated by a free parameter ($\beta$), but again a
unique value ($\beta \sim$ 0.1) was derived for it, independently on
the stellar mass \citep[see e.g.][]{st06}.

There is ample space to revise the above assumptions in the attempt
to explain the s-process abundances in OCs; however, as mentioned,
the revisions must preserve the interpretation of the abundances in
s-process-enriched AGB stars, which have already been successfully
achieved \citep{b+01,a+02}. As these last are known to have masses
$M$ $\gtrsim$ 1.5 \ms, we have a large freedom in choosing the
parameters for the \ctb pocket in stellar masses lower than this
limit. If born today (at solar or higher-than-solar metallicity)
such low mass stars would barely activate, during the AGB stages,
the third dredge-up (TDU) necessary to carry He-burning products to
the surface. However, the efficiency of TDU rapidly increases for
decreasing metallicity, so that at progressively lower values of
[Fe/H] we expect that stars of progressively lower mass can
contribute (albeit with a limited number of pulses and TDU
episodes). One of the scopes of this paper is to explore their
impact on the Galactic chemical evolution, especially for the recent
epochs in which OCs were formed.

We take general predictions for mass loss rates, TDU efficiency and
total number of pulses from the work by \citet{str03}. This paper
provides us with the general stellar parameters, which have been
often used in works where a global picture of AGB nucleosynthesis
was needed, like e.g. \citet{w+06}. The stellar models described in
\citet{str03} are also the same used in previous Galactic chemical
evolution calculations for the s process \citep{t+99,t+04}. By
adopting them we can verify the effects introduced exclusively by
the new proposal here advanced.

In the present work, we have modeled the formation of the \ctb
pocket, for stellar masses below 1.5\ms, by assuming that hydrogen
can be injected into the He-rich layers at TDU. The abundance of
protons $X_{\rm H}$ is assumed to exponentially decrease for
increasing depth (in mass) into the He zone. The maximum depth
reached was set to 4$\times$10$^{-3}$\ms, i.e. 4 times larger than
currently assumed \citep[cf][]{g+98}. At the restart of shell-H
burning, the upper part of this region becomes enriched in $^{14}$N,
a strong neutron filter hampering s-processing. The effective zone
where the nitrogen competition is negligible turns out to correspond
to a total mass of \ctb available for s-element production $M_{13} =
$ 1.2 $-$ 2 $\times 10^{-5}$\msb (depending on the details of the
proton profile assumed). This means that, over the depth of the
pocket (4$\times$10$^{-3}$\ms), we have an average {\it effective}
abundance by mass, $\bar X_{13}$, in the range $\simeq 3 - 5\times
10^{-3}$. For more massive AGB stars, assumptions very similar to
those discussed by \citet{t+04} were instead maintained, with an
average mass of \ctb burnt per cycle of 3.1$\times$10$^{-6}$\ms.

With the above prescriptions we computed the s-process {\it yields}
(i.e. the contributions in solar masses from the AGB stellar winds
to the ISM). The set of neutron-capture cross sections and of other
nuclear inputs was taken from the KADONIS database and from the
further recent improvements; see e.g. \citet{k+11} for a general
discussion.

Our stellar yields were then introduced into an improved version of
the Arcetri Galactic chemical evolution model, which is very similar
to the code used and updated by \citet{t+99,t+04}. For the present
work we have considerably improved the code. This revision included:
i) adopting a fine mass resolution (0.05\ms) for the contributing
AGB stars in the range of interest (1.25$ \lesssim M \lesssim$
1.5\ms); ii) using a detailed dependence on metallicity of the
stellar yields (providing them at seven different values of [Fe/H]);
and iii) estimating the Galactic enrichment using relatively small
time steps (25 Myr), in order to better describe the final phases of
the Galactic chemical enrichment. In particular, sampling properly
the change of the stellar yields with metallicity is of paramount
importance; in many previous models the stellar contributions were
estimated at few metallicity values (sometimes only two), but such a
course grid is not adequate for following s-element abundances. This
is clear, e.g., from \citet{b+99}, in particular their Figure 12,
where it is shown how, for any choice of the \ctb neutron source,
the stellar s-element yields do not increase monotonically with time
and/or metallicity, so that their trends must be sampled carefully.
Since s-processing is dominated by LMS contributors, also a high
resolution in time and stellar mass is important, as for them a
minimum variation in mass corresponds to large variations in the
evolutionary lifetimes, $\tau_{ev}$. These last were improved using
the computations by \citet{dom9}, which refer to the same FRANEC
code from which the stellar evolution scenario is taken.

For the rest, the technique adopted and the main Galactic-model
ingredients are very close to those of previous works on the
subject, so that the results can be immediately compared with the
ones by \citet{t+99,t+04}. In this respect we have adopted, for the
sake of consistency with previous works, the solar meteoritic
abundances from the compilation by \citet{ag89}. On the other hand,
for internal consistency, when referring to differential stellar
data, we have used the solar photospheric abundances derived in
Paper I for Y, Zr, La, and Ce and those from D'Orazi et al. (2009)
for Ba. Those abundances are indeed on the same scale as those of
open clusters, so that no systematic offset is introduced.

\section{The stellar yields}

The yields coming from very low mass stars in the model proposed
here are different than previously assumed, as the larger mass of
the \ctb reservoir now adopted affects them strongly. The changes
concern the total production factors in the winds and their trends
with metallicity and time. A peculiar characteristic of the cases
with a high \ctb mass to burn but with a small total stellar mass,
hence a limited number of pulses, is that the abundances in the He
shell can be sampled by dredge-up before an asymptotic composition
is reached in He layers. As a consequence, the s-element
distribution carried to the surface changes greatly from pulse to
pulse. In the first 3-4 TDU episodes mainly $ls$ nuclei are
contributed, while subsequently $hs$ elements become more important.
The final abundance pattern in the stellar wind is achieved through
the averaging operated by convective mixing over the continuously
changing composition that is dredged-up. In any case, no Population
I AGB star with $M < 1.5$\msb is found to experience more than 10-11
TDU episodes \citep[in accord with][]{b10}.

We note however that the predicted yields depend on the efficiency
of dredge-up, which is an uncertain parameter of any model. In the
parametrization by \citet{str03}, eq. (3), it is presented as a
function of the metallicity, of the H-exhausted mass ($M_{\rm H}$)
and of its increase during an interpulse period ($\delta$M$_{\rm
H}$). The values of these last two parameters depend crucially on
the rates for H-burning reactions, and in particular on that for
$^{14}$N(p,$\gamma$)$^{16}$O, the most critical one. The recent
changes in the recommended value for this rate \citep{ade}, not
included in \citet{str03}, certainly affect the TDU scenario: in
particular, small variations in $M_{\rm H}$ and $\delta$M$_{\rm H}$
easily lead to a change in the minimum stellar mass contributing at
Galactic disk metallicities, moving it inside the range 1.25 $-$
1.35\ms. Important effects on the amount of dredge-up are also
induced by the set of opacities used and by the treatment of
convection \citep[see e.g.][]{cr9}. These are the most typical
uncertainties that affect today the AGB inputs to Galactic chemical
evolution models of s elements. As a consequence, all results must
be taken with some caution; they can only provide a general guidance
and not precise quantitative conclusions.

\subsection{Stellar yields as a function of [Fe/H]}

\begin{figure}[t!]
\includegraphics[width=0.8\textwidth]{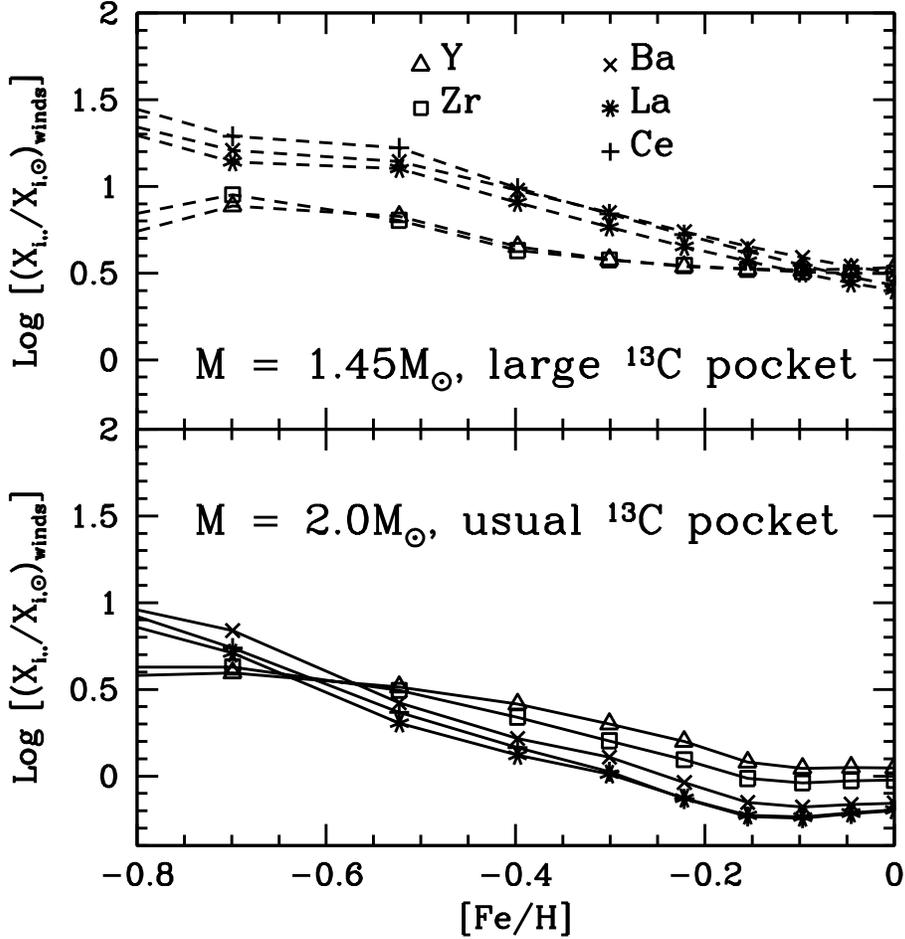}
\caption{\textit{Upper panel}: The s element contributions from a
stellar mass below 1.5\ms, expressed as production factors in the
stellar winds with respect to the solar composition, adopting the
extended \ctb pocket described in the text. \textit{Lower panel}:
the same contributions as obtained for higher masses with a standard
\ctb pocket.} \label{lifig8}
\end{figure}

For values of [Fe/H] relevant for the Galactic disk, our typical
yields are shown in Figure 1. The upper panel refers to our new
assumptions on the \ctb pocket for very low masses ($M <$ 1.5\ms),
using a \ctb abundance $X_{13}$ = 4 $\times$10$^{-3}$. For
comparison, the lower panel shows the yields for stars with $M
\gtrsim$ 1.5\ms, which are obtained with the older scenario and the
older (smaller) choice for \ctb mass burnt per cycle. The upper
panel, in particular, shows the situation for stars between 1.4 and
1.5\ms, which still have a rather efficient dredge-up for
metallicities up to the solar one. It shows that the stellar wind
composition, with the new \ctb pocket, is characterized by similar
production factors for $hs$ and $ls$ elements (as compared to the
solar distribution) only at a relatively high metallicity ([Fe/H]
$\gtrsim -$0.2). At lower [Fe/H] values the $hs$ are more
effectively produced. The distribution is remarkably different than
obtained for higher masses (lower panel). This is so both for the
absolute values and for the $[hs/ls]$ ratios. In particular, the
absolute yields at high metallicities remain high {\it only} for
lower mass stars, with the larger \ctb pocket, while they are
essentially negligible for higher masses, with the smaller, normally
assumed \ctb reservoir.

The relevance of the trends shown in Figure 1 can be understood with
reference to the age-metallicity relation of our model, illustrated
in Figure 2. It is derived a priori, from the reproduction of the
main species, like oxygen, the $\alpha$-rich elements and iron and
its behavior is typical in all modern Galactic evolution models. For
our purposes, Figure 2 shows two important things: i) the SFR has
remained rather high throughout the last 10 Gyr, after its peak in
earlier epochs; ii) during the last 7.5 Gyr of the Galactic history
the metallicity [Fe/H] has grown very little: over this whole, long
era, it has remained within $\pm$ 0.2 dex from solar. When
considered together, Figures 1 and 2 tell us that most of the stars
in the Galactic disk were born with high metallicity; moreover,
those having mass below 1.5\msb (dominating the Initial Mass
Function) generated, in their AGB phases, s-process yields close to
those at the extreme right of Figure 1 (upper panel). The larger
yields characterizing lower metallicity stars in Figure 1, with
element ratios very far from the solar distribution, played only a
marginal role in setting the composition of recent Galactic stellar
systems, because too few stars were born in those conditions. These
few stars could synthesize effectively only elements whose
overabundances in the stellar winds were extremely high. It is known
that this is the case with lead, for which low-metallicity AGB stars
provide the so-called {\it strong} s-process component \citep{g+98},
producing about 50\% of $^{208}$Pb.

\begin{figure}[h!!]
\begin{centering}
\includegraphics[width=0.8\textwidth]{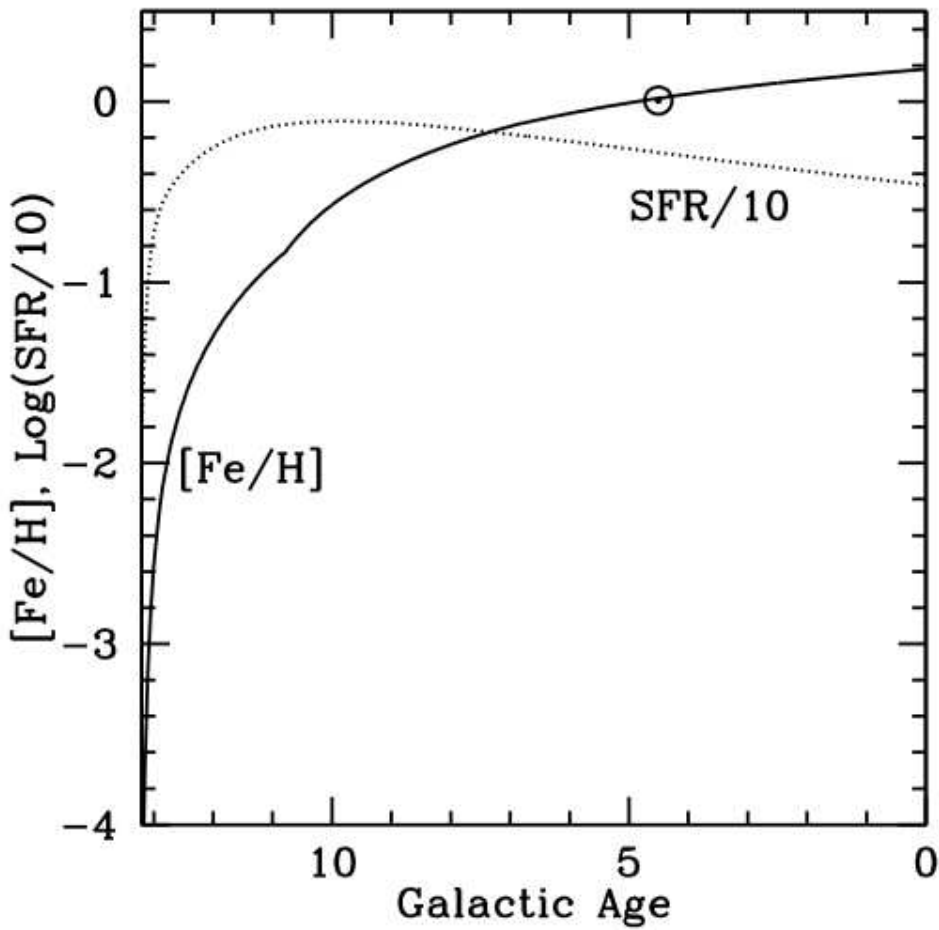}
\caption{The continuous curve shows the age-metallicity relation
obtained by our model, i.e. the history of the iron increase in time
produced by the Galactic evolution calculation. The dotted line
reports the Star Formation Rate (SFR), i.e. the mass going into
stars over the whole Galactic disk at each epoch, in units of
\ms/yr. The values of this function are divided by a factor of ten,
in order to fit it in the ordinate scale.}
\end{centering}
\end{figure}

\subsection{The Stellar yields as a function of time}

\begin{figure}[h!!]
\begin{centering}
\includegraphics[width=0.8\textwidth]{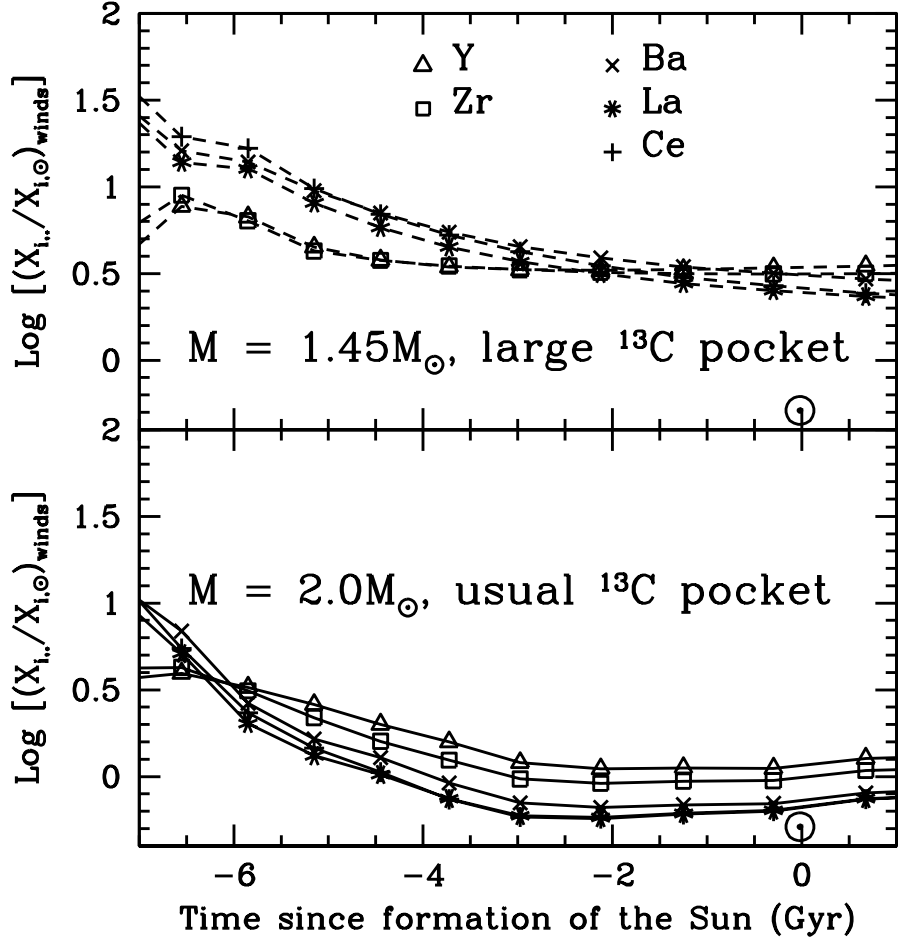}
\caption{The same enhancement factors in the stellar winds shown in
Figure 1, plotted as a function time. The zero of the scale
indicates the formation of the Sun. \textit{Upper panel}: during
more than 3 Gyr, before the solar formation, the production factors
of s elements in the winds of very low-mass AGB stars have remained
very similar. \textit{Lower panel}: more massive stars, with a
smaller \ctb pocket, could not play a significant role at high
metallicities, because of their negligible yields.} \label{fig4}
\end{centering}
\end{figure}

In order to make the above statements clearer we can actually
transform Figure 1 into a plot as a function of time, using the
age-metallicity relation of Figure 2 and setting the zero of the
scale at the moment of the solar formation. This plot is presented
in Figure 3. As one can see, the small region at the right side of
Figure 1 (for [Fe/H] values larger than about $-$0.2), where the
yields of $ls$ and $hs$ nuclei are about similar (within 0.2 dex),
is now shown to start typically 3 Gyr before the solar formation.
According to \citet{dom9}, such an interval allows for the complete
evolution of stars in the range $1.4 - 1.5$\ms, whose lifetime, in
this metallicity interval, ranges from 2.5 to 3 Gyr. As mentioned,
at the relevant [Fe/H] values these stars did have TDU; hence they
could affect the composition of the Sun and of stars born
subsequently with about equal production factors for $ls$ and $hs$
elements, thus clarifying how these two groups of nuclei can
increase their abundances in OCs, while maintaining their ratios
relatively constant. We emphasize that, as Figures 1 and 3 show, at
the times and metallicities mentioned, more massive stars had
essentially no role on s-processing, because their yields were far
too low.

The above figures make clear why the role of LMS in the chemical
evolution of s elements can change remarkably relative to previous
estimates, if a more extended \ctb pocket is assumed.

\subsection{The yields of a typical single AGB model \label{single}}
The $ls$ and $hs$ contributions to the Galactic enrichment derive
obviously from many stars of different mass and metallicity.
Nevertheless, we can identify a typical s-element nucleosynthesis
progenitor, providing yields for $ls$ and $hs$ elements whose mutual
ratios are very close to those of the subsequent Galactic average
(and of the solar distribution). With our large \ctb pocket, this
turns out to be an AGB star of 1.4$-$1.45\ms, with [Fe/H] $\sim
-0.1$. By normalizing to the abundances of s-only nuclei produced by
such a model star, we can infer s-process percentages in its winds
for the elements analyzed here. Using this estimate for explaining
the solar abundances we obtain 86\%, 84\%, 80\%, 85\%, 70\% and
76\%, for Sr, Y, Zr, Ba, La and Ce, respectively (see also later,
Table 1). From this, assuming that the complement to one of the
percentage abundances of the heavier elements (Ba, La and Ce)
derives from the r process, we infer r-process components of 15\%,
30\% and 24\% for Ba, La and Ce, respectively. These estimates are
slightly higher than those by \citet{b10}.

We now present the results on the abundance evolution of open
clusters induced by our new hypotheses and then briefly discuss the
consequences for the solar distribution.

\section{The resulting s-process composition of the Galactic disk}

\subsection{Interpreting the observations of open clusters}
In order to compute model expectations for the heavy elements in OCs
we must necessarily include in the calculations estimates for their
production in other astrophysical environments, outside the range of
LMS to which our models refer. As already mentioned in Subsection
\ref{single}, for Ba, La and Ce one can infer a rough estimate for
their r-process contributions from the yields of our model star best
reproducing the solar {\it main component}.

For lighter nuclei, the yields from IMS were estimated assuming that
no \ctb burning occurs in them: the neutrons derive only from the
activation of the \neanb reaction in thermal pulses. They induce a
secondary-like neutron-capture process, hence the yields vary as a
function of the metallicity. The total IMS production of s elements
is very uncertain, as these stars evolve almost entirely in the
infrared during the thermally-pulsing phase, so that observational
constraints are minimal and no trustable calibration of their mass
loss rates, suitable for their use in stellar models, exists. As a
consequence of this, the number of TDU episodes they undergo is
unknown. While, from nucleosynthesis models in thermal pulses, we
can derive the s-element distributions in their winds pulse after
pulse, not knowing the number of these last that can occur makes
their global contribution to the chemical evolution of the Galaxy a
rather free parameter, which must be calibrated ad-hoc. The \neanb
source is also activated during core-He and shell-C burning phases
of massive stars \citep[providing the so-called {\it weak
component}, see e.g.][]{bg85}. For estimating the corresponding
yields we modified previous estimates by \citet{r+93} following the
recent analysis by \citet{p+10}, performed with updated cross
sections for a solar-metallicity 25\msb star. The production in
stellar winds was then scaled at different metallicities assuming a
secondary-like behavior.

Despite the fact that the uncertainties affecting s processing in MS
and IMS are very large, they are not critical for our purposes, as
their joint contributions to Sr, Y and Zr do not account for more
than about 10\% of the total solar inventory for these species
\citep{s+09,b10}. Concerning the r-process (or, in general, the
primary) components of Sr, Y and Zr, in the presence of huge
uncertainties we estimated them to be around 5\%. Due to their low
values, these rather arbitrary choices are not critical in any
respect. Note that \citet{s+09} gave for them values between 5 and
11\%.

In order to compute the Galactic chemical evolution of the five
observed elements, we proceeded in the following way: i) we derived
metallicity-dependent stellar yields for the weak component in
massive stars and for the IMS (3-8\ms). ii) We then assumed that the
r-process components of the studied elements are produced at the
lowest mass end of type II Supernovae, as in \citet{r+99} and in
\citet{t+99}. iii) With the above ingredients, we followed the
chemical evolution of the Galaxy with our model, fixing the
uncertain value of the total mass contribution from IMS so that, at
the epoch of the Sun's formation and at the corresponding
Galacto-centric radius, the sum of all non-LMS contributions gives
the complement to one of our "best average" AGB model (see
Subsection \ref{single}). This means that the percentages form
astrophysical sources different from LMS must be of 16\% for Y, 20\%
for Zr, 15\% for Ba, 30\% for La and 24\% for Ce (see also Table 1,
column 4, Case C). iv) After this calibration, we introduced the
yields of LMS AGB stars, thus computing the total abundance of the
studied species as a function of time in the Galactic disk, up to
the recent eras when OCs were formed.

With the above assumptions, the Galactic chemical evolution model
produces, for the five measured elements, the evolutionary trends
shown in Figure 4, where the OC data are shown as empty circles;
each of them represents the average of the abundance over all the
stars analyzed in the cluster \citep[usually five; for details,
see][]{m+11}. The data for field stars are indicated with crosses of
the same size as the errorbar and are taken from the SAGA database.
In the left panels, adopting the common way of looking at the
chemical evolution of heavy elements, we plotted the observations as
a function of the iron abundance or metallicity, [Fe/H], while in
the right panels we showed them as a function of time. As already
mentioned in Section 1, the plots as a function of metallicity give
a confused pattern. This is so because the large scatter in the iron
abundances at any given age in the last few Gyr of Galactic life
does not make [Fe/H] a good evolutionary indicator for these stages.
This finding is in agreement with previous indications by
\citet{mash2}.

\begin{figure}[t!!]
\begin{centering}
{\includegraphics[width=0.78\textwidth]{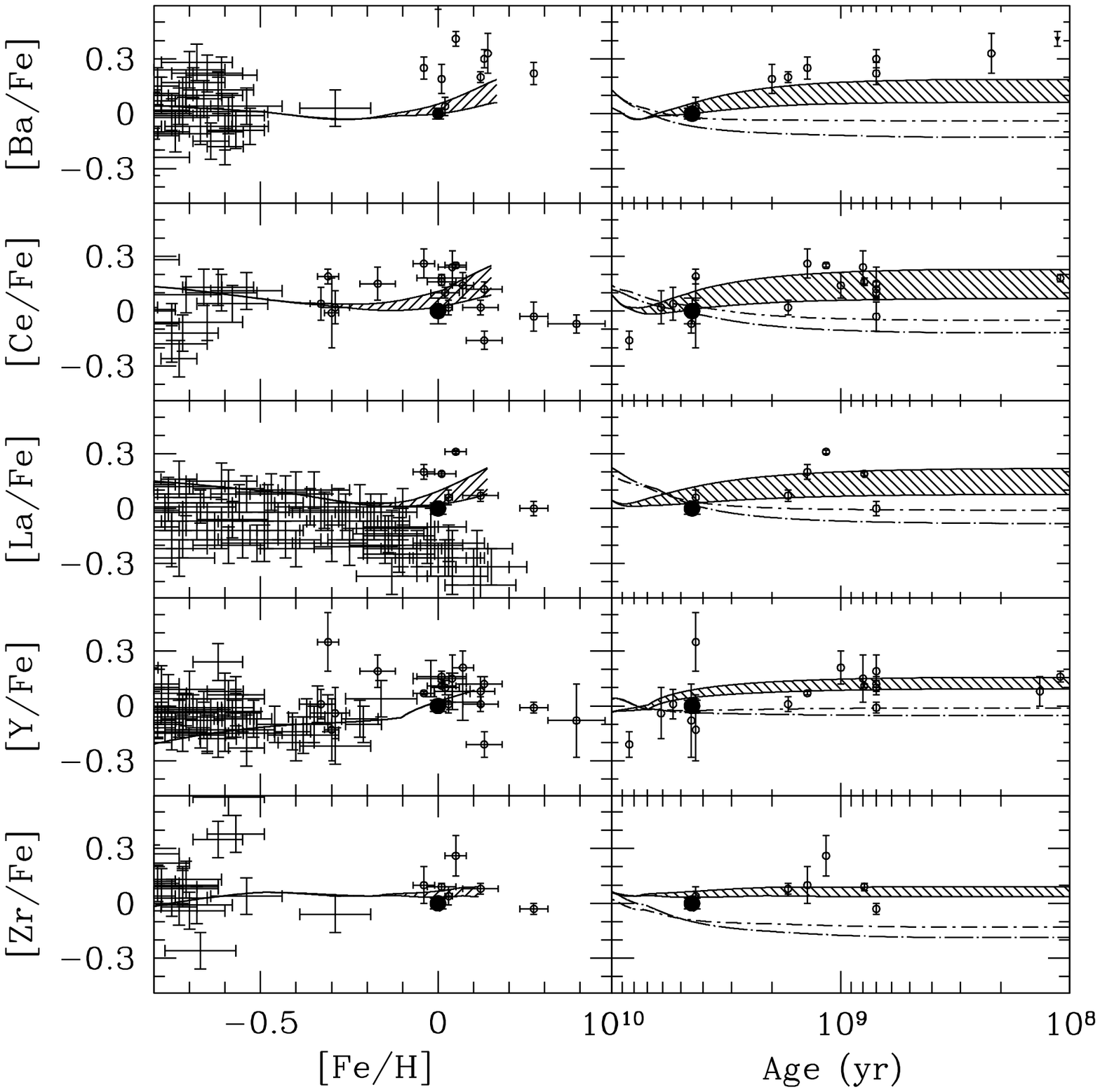} \caption{\footnotesize The Galactic enrichment in $ls$ and $hs$ elements from our chemical
evolution model, as compared to observed data. The solar
photospheric abundances are shown as heavy dots. The left panels
show the abundances as a function of [Fe/H]; the data for single
stars are from the SAGA database by \cite{saga}; their errorbar is
not always available, but certainly non-negligible. As an example,
the estimates deducted from Table 3 in \citet{sim} for La yield an
uncertainty of 0.1 dex. Even assuming this value (rather
optimistically) as typical for all the measured elements, as done in
the plot, we see that the emerging situation is very confused and a
definite trend for all the species studied does not emerge. Simply,
[Fe/H] is not an adequate parameter to describe the evolution of s
elements. The data for OCs are taken from Paper I and from
\citet{dor09}. Each open symbol represents a different cluster;
several among them have metallicities higher than solar. In the
right panels the abundances are instead plotted as a function of
time: here, only the clusters can be displayed (see text). While the
data are limited in number, they appear to show clear regularity,
without a wide scatter. The shaded area, indicating our model
predictions, reproduces the  solar and clusters' measurements within
the uncertainties of the observations and of the models (see text
for explanations). In the right panels, dash-dotted lines show the
trends obtained without the extension of the \ctb pocket suggested
in this paper, either including (short dashes) or excluding (long
dashes) masses below 1.5\ms.}}
\end{centering}
\label{lifig9}%
\end{figure}

Only when the evolution in time is properly reconstructed (right
panels) the recent growth of s-elements is unveiled. This is however
easily done only for the clusters. Their age is known from a variety
of methods: our specific choices, discussed in Paper I, were taken
from \citet{msrg}. For clusters older than 0.5 Gyr, they made use of
the morphological age indicator $\delta V$ \citep{P94}, as
calibrated by \citet{sal04}; for younger clusters they used the most
recent age estimates available, like those obtained by the lithium
depletion boundary method. For field stars, similar criteria do not
exist and a precise value of their age is usually not available, so
that observations referring to them (present in the left panel)
cannot be plotted as a function of time.

The s-element abundances in OCs displayed in Figure 4 were measured
differentially with respect to the solar photosphere, using the same
method of analysis and the same line list (see. Paper I
for details). For this reason, the data were normalized to the
photospheric solar abundances. These are, in some cases, slightly
different from the meteoritic ones (that were used in our "average
model" of the solar main component and in deriving the s-process and
r-process contributions to the chemical evolution). This explains
why, for some elements, the model curves pass above the solar
position: it is only an appearance, deriving from the normalization.

In each panel of Figure 4, the shaded areas represent predictions
from our calculations and aim at giving an idea of the model
sensitivity to the most crucial parameters, i.e. the abundance of
\ctb burnt and the lower mass limit for TDU operation. The shaded
ranges take into account both a variation by $\pm$25\% around the
average value of our \ctb abundance, $X_{13} = (4 \pm 1)
\times$10$^{-3}$, and the uncertainty on the lowest mass
contributing to the process, as discussed in Section 3. In the right
panels, two more evolutionary curves are plotted: the long
dash-dotted line shows the evolutionary trend obtained by excluding
the masses below 1.5\ms; the short dash-dotted one represents
instead a case in which they are included, but using for them the
same extension of the \ctb pocket adopted for higher mass stars. It
should be clear, from the right panels of the figure, that the extra
production we assumed from low mass AGB stars offers a means to
account for the recent observations of young OCs.

It is relevant to mention that the spectroscopic observations in
clusters, which are our original constraints, do not offer us
indications with an accuracy better than about $\pm$ 0.1 dex; within
these limits we can say that the chemical evolution model, despite
all its uncertainties, suffices to account for the data, namely for
a global increase in the abundances that nevertheless maintains a
solar-like distribution. This is the main point we want to stress:
s-element abundances in Population I stars can be explained, within
their uncertainties, by suitably fixing the parameters controlling
the neutron source in low mass stars, without the need of further
ingredients.

\subsection{LMS yields and the solar distribution of heavy elements}
As a further test, we ran a few cases in which we included in our
Galactic model only the yields of Figures 1 and 3, excluding any
further, uncertain contribution from other sources (IMS, MS and the
r process). This allowed us to derive an estimate of the role
exerted solely by LMS in determining the composition in heavy
elements of the recent Galaxy and of the Sun. This role is
illustrated in Table 1. There, we put in evidence the uncertainties
in the models mentioned before, by showing two possibilities (called
cases A and B, presented in columns 2 and 3), where we include (or,
respectively, exclude) in the evolutionary calculations the lowest
mass bin from 1.25 to 1.3\msb (see the discussion in Section 3 for this).

As a comparison, we also show as model C, presented in column 4, the
s-process percentages obtained in the yields of our
already-mentioned single AGB star, of 1.4$-$1.45\msb and with [Fe/H]
$\simeq -$0.1, best approximating the Galactic average and the solar
main component.

\begin{table}[h]
\footnotesize{ \centerline{Table 1} \centerline{Percentage
Contributions to Solar Heavy Elements from LMS} \centering{
\begin{tabular}{|l|c|c|c|c|}
\hline Element & This work (A) & This work (B) & This work (C) & Literature\\
\hline
Strontium & 90 & 89 & 86 & 55$^{(1)}$   \\
Yttrium   & 81 & 80 & 84 & 62$^{(1)}$   \\
Zirconium & 80 & 78 & 80 & 55$^{(2)}$   \\
Barium    & 89 & 86 & 85 & 84$^{(2)}$   \\
Lanthanum & 72 & 70 & 70 & 70$^{(2)}$   \\
Cerium    & 79 & 78 & 76 & 81$^{(2)}$   \\
\hline
\end{tabular}

Cases (A), (B): chemical yields from LMS only, with/without the lowest mass bin \\
Case (C): composition of the winds of a single AGB star, 1.45\ms, [Fe/H] = $-$0.1\\
$^1$ \citet{t+04}; $^2$ \citet{b10}, single star\\
}}
\end{table}

Despite the necessary cautions related to the models, the exercise
shown in Table 1 (cases A and B) makes clear that pure s processing
in LMS with a suitable choice of the parameters for the \ctb pocket
can account for typically 80\% (sometimes more) of all the solar
s-element abundances. This is so for both $ls$ and $hs$ nuclei,
without distinction. As a comparison, in Table 1 (column 5) we
report the percentages of $ls$ elements attributed to LMS by
\citet{t+04}, without the \ctb pocket enhancement. These authors
underlined that 20-30\% of the abundances were not accounted for,
even after including the contributions from IMS, massive stars and
the r-process. In our new scenario, depicted in Table 1 (columns 2
and 3) the LMS production of the elements considered is increased
just by the amount that was missing (20 - 30\%), so that the solar
abundances of Sr, Y and Zr are now satisfactorily reproduced. We
cannot therefore confirm the need, suggested by other authors, for
an unknown primary process, sometimes called $LEPP$, or {\it Lighter
Element Primary Process}, contributing to these elements \citep[see
e.g][for details on this proposal.]{t+04}


\begin{table}[t!]
\footnotesize{ \centerline{Table 2} \centerline{Percentage
contribution to solar s-only nuclei} \centering{
\begin{tabular}{|l|c|c|c|}
\hline
Isotope & Case A & Case B & Arlandini et al.(1999) \\
\hline
   $^{100}$Ru & 95 & 93  & 95 \\
   $^{110}$Cd & 97 & 95  & 97 \\
   $^{124}$Te & 94 & 92  & 91 \\
\hline
\end{tabular}

}}
\end{table}

In order to substantiate our statements about s processing, Table 2
reports our LMS contributions to a few s-only nuclei of the solar
distribution with $A < $ 130. The results are presented for the
cases A and B defined above. Again, considering the 10\% errorbar in
solar abundances, plus all the model uncertainties, it seems that
the solar concentrations of s-only nuclei can be rather
satisfactorily explained. In particular, Table 2 shows that our
stellar yields, integrated through the chemical evolution model up
to the formation of the Sun, can account for essentially the same
percentages that could be previously obtained only from average
single-star models \citep[see][]{a+99}.

Note that we consider Tables 1 and 2 only as indications that the
solar s-process elements can indeed be explained by pure
s-processing. It would be completely outside our scopes to search
for a really detailed fit to the solar abundances through a full
chemical evolution procedure.

\section{Discussion and Conclusions}
In the previous sections we presented a tentative interpretation of
the increase in heavy element abundances after the Sun's formation,
as revealed by recent spectroscopic observations of young open
clusters. The increase affects elements that are predominantly
produced by slow neutron captures and we showed that their trend in
recent stellar systems can be understood as a result of the
contributions from very low mass, long-living stars ($M <$ 1.5\ms,
$\tau_{ev} \gtrsim$ 2 Gyr), if they activate the main neutron source
more effectively than previously suspected.

A consequence of the above suggestion is that also for the Sun the
s-process main component can be naturally accounted for, without the
need of any extra contribution. An additional conclusion that might
be obtained from our procedure concerns the behavior of
predominantly-s elements at low metallicity, in comparison to those
that are mainly produced by the r-process, like e.g. Eu. In our
cases A and B of Table 1 we found, for the s-process component of
Eu, the values of 7.5 and 6.5\%, respectively. From this, and making
the guess that the complement to one of the abundances for Ba and Eu
is due to the r-process, one might deduce a ratio [Ba/Eu]$_r$
ranging from $-$0.8 to $-0.9$. This is roughly compatible with the
average values measured in r-process-enriched metal-poor stars,
excluding the largely scattered measurements at [Fe/H] $\lesssim
-2.5$ \citep{saga,sne}. However, due to the mentioned uncertainties,
we believe that such an extrapolation to the r-process is not really
robust (although it is usually accepted in the literature). The
stellar and Galactic parameters cannot simply be fixed accurately
enough to yield a real confidence in such estimates.

Understanding how our scenario is produced is now easy, in the light
of the results shown in Figures 1 to 3. What is crucial in this
context is the evolution {\it in time}: most parent AGB stars were
born in the last 7.5 Gyr, when the metallicity is high and suitable,
with the larger \ctb pocket we suggest, to produce $ls$ and $hs$
elements in similar proportions (similar to those of case C in Table
1).

We notice that, regardless of the specific model we propose here,
the same observational data offer a picture of the heavy element
evolution in the Galactic disk that is incompatible with the
hypothesis that fast primary processes strongly contribute to $ls$
elements. Indeed, the fact that the abundances of Y, Zr, Ba, La and
Ce grow with respect to iron (and do so very late in the Galactic
history) excludes a primary process, the products of which would
follow the trend of iron itself. Moreover, the suggestion that a
supplementary process modifies selectively the lighter
neutron-capture nuclei is contradicted by the fact that in young
stellar systems Y and Zr show the same trend as Ba, La and Ce.
Furthermore, below $A$ = 130 there are a few s-only isotopes,
shielded from fast decays, whose production in adequate
concentrations requires necessarily the action of slow neutron
captures, not of fast processes.

It is known that the $LEPP$ has been invoked also (and mainly) for
extremely metal poor stars. Here the classical idea that fast
neutron captures (those of the r process) account for the whole
abundances of heavy elements \citep{trur} was subsequently
integrated and modified by new observations \citep{sne}. First
suggestions on the existence of different stellar sites for
producing r-process elements of different atomic mass number were
advanced from an analysis of the record of live radioactive nuclei
in the Early Solar Nebula \citep{wbg}. A large observational basis
on metal-poor and extremely metal-poor stars now supports the fact
that, while above Te-Xe the r-process does appear to be rather
universal, the abundances of lower mass elements are more disperse
and more difficult to interpret \citep{ao05,saga,roed}. It was
noticed that certain heavy nuclei appear to be produced in
progenitors forming very little iron \citep{qw07}; it was also
suggested that elements at the neutron magic number $N$ = 50 might
derive from a combination of charged-particle reactions and
neutron-captures in high-entropy neutrino winds associated to
core-collapse supernovae \citep{qw07,wq09,roed,th10}. It is in any
case beyond doubts that lighter and heavier r-process nuclei should
come from different sources or from different superpositions of
astrophysical conditions \citep[see e.g.][]{far,th11}.

The above confused situation is today the object of a lively debate.
Unfortunately, the data and models discussed in this paper cannot
offer any clue for the understanding of extremely metal-poor stars.
We notice that, also in that case, attempts at explaining the
complex observational scenario with suitable combinations of
existing mechanisms, without invoking new ad hoc processes, already
exist \citep{qw08,wq09,far,bhq}.

We conclude by mentioning that the extended \ctb pocket assumed in
this paper would require the existence of very efficient mixing
episodes in evolved stars of mass below 1.5\ms. A similar conclusion
was already inferred from very different constraints
\citep{p11a,p11b}; in particular, it seems to be necessary for
explaining the isotopic admixture of oxygen in presolar oxide
grains. The transport mechanisms most commonly-adopted so far
(rotationally-driven shear, or thermohaline mixing) would not
suffice \citep{p11a}. From this point of view, the apparent need of
forming extended \ctb reservoirs in the short time interval of a TDU
episode, might be another indication that other mixing processes
must be at play. Phenomena of this kind were recently suggested to
be driven by magnetic buoyancy \citep{b+07,n+08}; this hypothesis
and other mechanisms must now be examined in order to see if it can
offer a general interpretation of non-convective mixing in evolved
stars.

{\bf Acknowledgments}. We are strongly indebted to the referee, G.J.
Wasserburg, for his constructive criticisms, which greatly helped in
shaping the paper as it appears now. Many colleagues provided us
with their advise; this is in particular true for C. Abia, A.
Chieffi, S. Cristallo, F. K\"appeler, O. Straniero, C. Travaglio.

\end{document}